\documentclass[runningheads]{llncs}

\usepackage{graphicx}
\usepackage{float}
\usepackage{soul}
\usepackage{float}
\usepackage{lipsum}
\usepackage[hidelinks]{hyperref}
\usepackage[utf8]{inputenc}
\usepackage[small]{caption}
\usepackage{amsmath}
\usepackage{cleveref}
\usepackage{graphicx}
\usepackage{booktabs}
\usepackage{csquotes}
\usepackage{dblfloatfix}
\usepackage[hyphenbreaks]{breakurl}

\begin{document}
\title{Combining ID's, Attributes, and Policies in Hyperledger Fabric}
%Combining Multiple ID's, Attributes, and Policies to Provide Secure Access Control within Hyperledger Fabric Blockchain Networks

\author{Daan Gordijn\inst{1} \and
Roland Kromes\inst{1} \and
Thanassis Giannetsos\inst{2} \and
Kaitai Liang\inst{1}}

\authorrunning{D. Gordijn et al.}
\institute{Cyber Security Group, Delft University of Technology, The Netherlands 
\email{D.A.Gordijn@student.tudelft.nl, R.G.Kromes@tudelft.nl,
Kaitai.Liang@tudelft.nl}
\and Ubitech Ltd, Digital Security \& Trusted Computing Group, Athens, Greece
\email{agiannetsos@ubitech.eu}
}

\maketitle              % typeset the header of the contribution
\begin{abstract}

\end{abstract}

\begin{abstract}

This work aims to provide a more secure access control in Hyperledger Fabric blockchain by combining multiple ID’s, attributes, and policies with the components that regulate access control.  The access control system currently used by Hyperledger Fabric is first completely analyzed. Next, a new implementation is proposed that builds upon the existing solution but provides users and developers with easier ways to make access control decisions based on combinations of multiple ID's, attributes, and policies. Our proposed implementation encapsulates the Fabric CA client to facilitate attribute addition and simplify the process of registering and enrolling a newly created certificate (corresponding to a new user). This research, concludes that it is possible to combine multiple ID's, attributes, and policies with the help of Hyperledger Fabric's smart contract technology. Furthermore, it could be seen that the performance impact for real-world applications is negligible compared to the insecure case of always providing access to a resource without performing access control.

\end{abstract}

\section{Introduction}
Ever since the anonymous Satoshi Nakamoto published his Bitcoin white paper \cite{NAKAMOTO2008} in 2008, blockchain has become one of the most disruptive technologies in the computer science industry. In recent years, many other innovative blockchain technologies have been developed \cite{number-of-blockchains}, which are becoming increasingly more popular. 

While Bitcoin was created to provide a digital alternative to traditional, bank-controlled currencies \cite{bitcoin-fiat-currencies}, many of these newer blockchain technologies are designed to provide a platform for building and deploying decentralized applications through the use of smart contracts\footnote{A digital contract written into code that is stored and automatically executed on the nodes of a distributed blockchain network \cite{smart-contracts}}. By implementing their business logic within these smart contracts, decentralized applications can automatically execute any transaction without human intervention, making them completely independent and decentralized \cite{TAPSCOTT2016}. Due to the many benefits of decentralized applications \cite{CAI2018}, the adoption of blockchain technologies has recently expanded to many non-financial applications such as “healthcare, supply chain management, market monitoring, smart energy, and copyright protection” \cite{XU2019}.

Most of these traditional blockchain technologies, such as Bitcoin, Ethereum, and Cardano, are so-called “permissionless” blockchain technologies. This type of blockchain technology, however, has many privacy issues when it is being used in the context of enterprise-level applications, as described in \cite{PENG2020}. Many alternative, so-called “permissioned” blockchain technologies have been proposed to solve the issues, the most promising of which is Hyperledger Fabric \cite{hyperledger-documentation}. Through the use of innovative concepts such as channels, policies, identities, and Membership Service Providers, Hyperledger Fabric can determine the identity of participants, perform access control based on these identities, and ensure the privacy of transactions and smart contracts.

As with many technologies, the increase in popularity of blockchain technologies also drives an increase in security threats and attacks. One of the major issues that many blockchain technologies, including Hyperledger Fabric, currently have is providing secure access control to the distributed ledger and smart contracts. Hyperledger Fabric partially addresses this issue by only granting network access upon submission of a valid X.509 certificate \cite{x509-certificate}, issued and approved by a trusted Certificate Authority. However, this type of ID-based access control is not scalable for larger organizations. This paper will therefore investigate how secure access control in Hyperledger Fabric can be improved, in particular by looking into solutions that can combine these ID's with attributes and policies.

This paper aims to study how secure access control in Hyperledger Fabric can be guaranteed by combining several IDs, attributes and policies with the components that regulate access control.
access control. 

The study also highlights the access control components that currently interact within Hyperledger Fabric. This work also provides a an implementation for combining  multiple ID's, attributes, and policies be combined within Hyperledger Fabric, and analyzes  the performance impact of ID-, attribute-, and policy-based access control. 

%The research question that has been formulated for this study is “How can secure access control in Hyperledger Fabric be guaranteed by combining multiple ID's, attributes, and policies with the components that regulate access control?”. In order to systematically answer this research question, the following sub-questions will be addressed:

%\begin{itemize}

%    \item How are the components for access control currently interacting within Hyperledger Fabric?
%    \item How can multiple ID's, attributes, and policies be combined within Hyperledger Fabric?
%    \item What is the performance impact of ID-, attribute-, and policy-based access control within Hyperledger Fabric?
%\end{itemize}

This paper is structured in the following manner. First, Section \ref{sec:RelatedWork} will provide a summary of the most relevant work that currently exists in literature. Next, Section \ref{sect:Contribution} will provide an overview of the contributions made by this research. Then, Section \ref{sec:Background} will provide a background on the current access control system of Hyperledger Fabric, while Section \ref{sec:ProposedImplementation} will present the proposed system model that has been implemented as part of this research. Subsequently, Section \ref{sec:Results} will provide an overview and analysis of the results that have been obtained during the research, while Section \ref{sec:Discussion} will provide a brief discussion. Finally, Section \ref{sec:Conclusion} will present the main conclusions of this research.

\section{Related Work}
\label{sec:RelatedWork}
Research into secure access control in various blockchain technologies, including Hyperledger Fabric, has been conducted in multiple papers. Many of these studies are performed in the context of exploring the integration of blockchain technologies with the Internet of Things (IoT), as blockchain is currently seen as the most promising technique for providing secure access control to IoT devices \cite{DING2019}.

In \cite{ROUHANI2019}, a summary of the major problems of modern access control systems is presented, together with an explanation of how these problems can potentially be solved using blockchain technologies. Furthermore, this paper provides an overview of existing access control studies and describes the current challenges of blockchain-based access control.

In \cite{DING2019}, an attribute-based access control scheme for Internet of Things devices is proposed by employing blockchain technology to keep track of the distribution of the attributes. Next, \cite{SONG2020} proposes a different scheme that is built upon various smart contracts and so-called “functional modules”, which are jointly responsible for managing attribute information and making access control decisions. Finally, \cite{YUTAKA2019} proposes yet another access control scheme which is implemented and deployed using the smart contract technology of the Ethereum blockchain network. 

While the papers discussed so so far describe general blockchain-based access control systems, other papers make specific use of the Hyperledger Fabric blockchain technology. First, \cite{IFTEKAR2021}, \cite{YANG2021}, and \cite{ISLAM2019} explore basic access control scenarios for IoT devices in Hyperledger Fabric. Next, \cite{ZHAO2022} combines the Hyperledger Fabric blockchain technology with the InterPlanetary File System (IPFS) \cite{ipfs-documentation}, allowing IoT devices to easily store documents on a distributed file system and store the hashes of these documents on the blockchain ledger. Finally, \cite{BANDARA2021} proposes a multi-layered and multi-model access control system in the context of an agricultural supply chain system that runs on Hyperledger Fabric. 

While research into secure access control in Hyperledger Fabric and other blockchain technologies certainly exists, no study into combining multiple ID's, attributes, and policies during the decision-making process has been conducted. To fill in this gap, this paper will propose a new access control scheme that combines these ID's, attributes, and policies within a single smart contract deployed to a Hyperledger Fabric network. For consistency, this research paper will consider the scenario where an IoT device wants to store a document on IPFS and subsequently save the returned document hash on the blockchain network, as also used in \cite{ZHAO2022}.

\section{Contribution}
\label{sect:Contribution}
%A new design to provide secure access control in Hyperledger Fabric was proposed, which was then implemented and analyzed during the \textit{implementation} part using a smart contract.
%The research conducted for this paper consisted of two main parts. First, a \textit{literature research} was conducted to study the existing solutions for secure access control, both within and outside the context of Hyperledger Fabric. Second, a new design to provide secure access control in Hyperledger Fabric was proposed, which was then implemented and analyzed during the \textit{implementation} part using a smart contract.

%\subsection{Literature Research}
%During the literature research, Google Scholar was used to find existing literature about secure access control in Hyperledger Fabric and other blockchain technologies. The exact search queries that have been used during this Google Scholar search can be found in Appendix 1. In addition to Google Scholar, the snowball method and citation searching method \cite{tulib-search-plan} have been applied to retrieve additional literature that was not listed in the initial search results. A full summary of the most important literature gathered during the literature research phase can be found in Section 2.

%\subsection{Implementation}
Using the existing literature from the previous phase, a new design was proposed to provide secure access control in Hyperledger Fabric. As stated in the research question, this design had to combine multiple ID's, attributes, and policies in the decision-making process. Subsequently, during the implementation phase, the design was implemented using a smart contract and deployed to a local Hyperledger Fabric test network, which was set up using the official tutorial \cite{hyperledger-using-test-network}.

Hyperledger Fabric currently supports three programming languages for the development of smart contracts and client applications: Go, Java, and NodeJS \cite{hyperledger-introduction}. For each language, several SDK's are available \cite{hyperledger-sdks} that help make the implementation of smart contracts 
and client applications easier. For this particular research project, NodeJS with TypeScript has been selected as the toolchain for the implementation phase, as this language is very easy to learn and understand.

The complete repository that contains a basic test network together with the smart contracts and sample applications that have been implemented during this research project is
available on GitHub\footnote{\label{github-repository}\url{https://github.com/daangordijn/Fabric-Access-Control}}. The \texttt{README} stored in this repository also includes a small tutorial, as well as a complete overview of the required tools and their recommended versions.

\section{Background}
\label{sec:Background}
This section discusses the current approach to secure access control in Hyperledger Fabric. This section begins with a brief introduction to Hyperledger Fabric and secure access control in general, and subsequently discusses the main components and methodologies that Hyperledger Fabric currently uses to provide secure access control.

\subsection{Hyperledger Fabric}
Hyperledger Fabric is an “open-source enterprise-grade permissioned distributed ledger technology (DLT) platform, designed for use in enterprise contexts” \cite{hyperledger-introduction}. While many well-established blockchain platforms such as Bitcoin and Ethereum are currently being modified to be used in enterprise-grade applications, Hyperledger Fabric has been built around enterprise applications from the beginning. First, Hyperledger Fabric is highly modular, which allows core parts of the blockchain network to be customized. Second, Hyperledger Fabric has support for writing smart contracts in general-purpose languages, including Go, Java, and NodeJS, while most other blockchain technologies require developers to learn new languages, such as Vyper or Solidity in the case of Ethereum \cite{ethereum-languages}. Finally, Hyperledger Fabric is permissioned, which means that the identity of all participants of the network is known and can therefore be verified using access control systems, allowing organizations to establish trust. 

Each node in the network maintains a local Membership Service Provider (MSP). These service providers store all X.509 certificates that have been issued by the Certificate Authorities of their corresponding organizations, which are then used by network nodes to map X.509 identities to internal roles. Together with the Certificate Authorities, these providers are therefore responsible for providing the initial layer of identity-based access control.

\subsection{Secure Access Control}
Access control is “a security technique that regulates who or what can view or use resources in a computing environment” \cite{access-control}. Different types of access control exist, including Identity-Based Access Control (IBAC), Role-Based Access Control (RBAC), and Attribute-Based Access Control (ABAC) \cite{DAIBOUNI2016}. While older, established blockchain technologies such as Bitcoin and Ethereum are non-permissioned and therefore do not implement these types of access control systems, Hyperledger Fabric is a permissioned blockchain technology, which enforces it to perform access control.

Currently, Hyperledger Fabric employs multiple layers of access control to provide security and privacy within the blockchain network. First, at the most basic level, Hyperledger Fabric uses a simple identity-based access control system, which prevents unauthorized entities from accessing anything on the blockchain network. This layer is explained in more detail in Sections 4.3 and 4.4 since the purpose of this paper is to extend this simple system to a more complex attribute-based access control system. Second, at an organizational level, Hyperledger Fabric can restrict access to smart contracts and the ledger through the use of channels, as described in Section 4.1. By only granting individual organizations access to the minimal required subset of channels, the privacy of smart contracts and ledger states can be preserved.

\subsection{Certificate Authorities (CAs)}
A Certificate Authority is an “organization that acts to validate the identities of entities and bind them to cryptographic keys through the issuance of electronic documents known as digital certificates” \cite{what-is-a-ca}. Hyperledger Fabric provides a special implementation, called the “Fabric Certificate Authority” or “Fabric CA” in short, which can be used to create and sign these digital certificates using the international X.509 standard \cite{x509-standard}. Fabric CA consists of both a client-side and server-side command line interface (CLI), called \texttt{fabric-ca-client} and \texttt{fabric-ca-server}, respectively. Fabric CA provides many features including “registration of identities, issuance of enrollment certificates, and certificate renewal and revocation” \cite{hyperledger-fabric-ca}. 

When an administrator wants to enroll a new identity, Fabric CA will generate a key-value pair that consists of a private key and a public key. Together with the parameters provided by the administrator, a Certificate Signing Request (CSR) will be created, which is then processed by Fabric CA. 

In Section 4.5, this process of registering and enrolling a new identity with the Fabric CA server is visualized. This section will also describe a new command line interface (CLI) that has been implemented as part of this study and makes the creation of new identities much easier.

\subsection{Membership Service Providers (MSPs)}
A Membership Service Provider is a component within Hyperledger Fabric that can be used by participants of the blockchain network to prove their identity to other participants of this network. When a user wants to start interacting with a Hyperledger Fabric blockchain network, it needs to create a key pair, which consists of a public key and a private key, which is needed to prove its identity to the rest of the network. Next, this public key must be included in a Certificate Signing Request (CSR), which is then submitted to a Certificate Authority and used to issue a new X.509 certificate. While X.509 certificates, including public keys, can be shared publicly, private keys must always be kept secret to comply with the principles of Public-Key Infrastructure (PKI) \cite{public-key-infrastructure}.

When a participant of the blockchain network now wants to submit a transaction, it needs to create a transaction proposal and sign this proposal using its private key. All nodes on the blockchain network are then able to verify this transaction proposal using the public X.509 certificate of this participant since it is stored inside the Membership Service Providers. Because of this, Membership Service Providers can establish trust on the permissioned blockchain network, without the need of sharing private keys.

\subsection{Generating Certificates}
\begin{figure}[h]
    \centering
    \includegraphics[width=9cm, height=7cm]{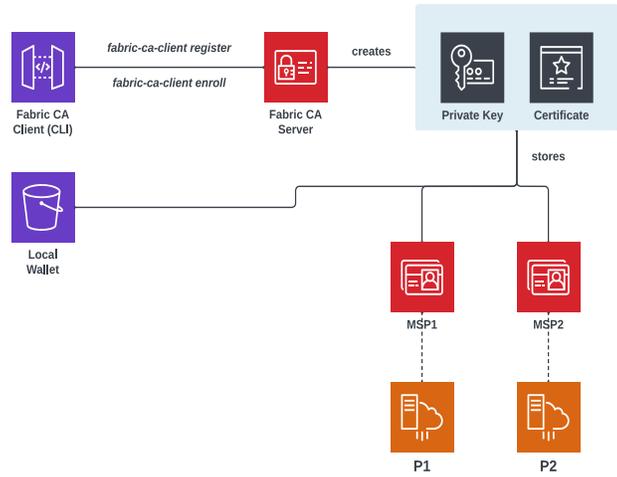}
    \caption{Current process of enrolling a new identity within a Hyperledger Fabric network. The \texttt{fabric-ca-client} CLI is used to run the \texttt{register} and \texttt{enroll} commands, respectively. Then, the resulting X.509 certificate is stored on a set of peer nodes, while both the X.509 certificate and the corresponding private key are stored in the user's local file system wallet.}
    \label{fabric-ca-msp}
\end{figure}

In Figure \ref{fabric-ca-msp}, a simplified version of the process of generating X.509 certificates using Fabric CA is visualized. As can be seen, the Fabric CA Client has to invoke the Fabric CA Server using two commands, \texttt{fabric-ca-client register} and \texttt{fabric-ca-client enroll} \cite{hyperledger-fabric-ca}. By doing this, the server will generate a private key, a public key, and a corresponding signed X.509 certificate. This certificate is then automatically stored in the Membership Service Providers that are located on various nodes inside the blockchain network.

While this process of generating X.509 certificates for Hyperledger Fabric is not overly complicated, it can become cumbersome to run multiple commands with many different flags to just create one certificate. Therefore, as part of this research paper, a wrapper around the \texttt{fabric-ca-client} was created. This tool, called \texttt{certgen}, is publicly available in the GitHub repository\cref{github-repository}, together with a small tutorial on how to interact with it. The \texttt{certgen} tool internally uses the \texttt{fabric-ca-client} commands and has the advantage that it can automatically populate a local file system wallet with the correct files which are required to connect a client application to the blockchain network. In addition, since this tool is highly interactive, it makes it much easier for administrators to add attributes to the certificate. More about the importance of setting attributes within X.509 certificates will be explained in Section 5. 

\section{Proposed Implementation}
\label{sec:ProposedImplementation}
This section discusses the proposed implementation that improves the current implementation of secure access control in Hyperledger Fabric, introduced in Section 4. This section begins with a brief discussion of how to independently combine multiple ID's, attributes, and policies, and subsequently presents the final design incorporating these components.

\subsection{Combining Attributes}
\begin{figure}[H]
    \centering
    \includegraphics[width=8cm, height=5cm]{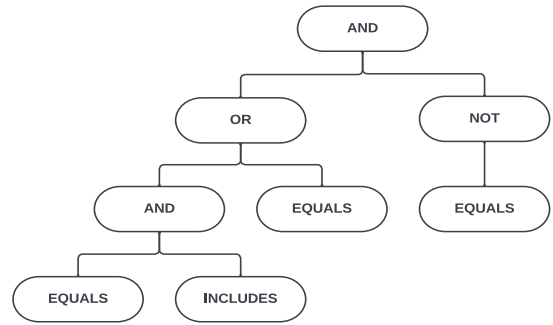}
    \caption{Combining multiple attributes. The \texttt{EQUALS} and \texttt{INCLUDES} operators validate whether a specified attribute equals or includes a certain value, respectively. The \texttt{AND}, \texttt{OR}, and \texttt{NOT} boolean operators can be then be applied to combine or negate these individual attribute checks, allowing the client to create complex policies.} 
    \label{combining-attributes}
\end{figure}

In Hyperledger Fabric, every X.509 certificate issued by Fabric CA \cite{hyperledger-fabric-ca} can have attributes. These attributes can be used during access control to determine whether a client should be given access, or not. To allow for more complex access control decisions, multiple attributes can be combined into so-called “policies”, which are visualized in Figure \ref{combining-attributes}. 

For this study, the following boolean operators have been selected that can be used for building access control policies:

\begin{itemize}
    \item \texttt{EQUALS}: Checks whether an attribute is present on the certificate, and whether it is equal to the provided value. 
    \item \texttt{INCLUDES}: Checks whether an attribute is present on the certificate, and whether it includes the provided value. This operator can be used when the specified attribute on the certificate has a comma-separated list of strings as its value, which must include a particular value. 
    \item \texttt{AND}: Logical operator that combines two or more operator trees. This operator returns \texttt{true} if and only if all operator trees combined by this operator evaluate to \texttt{true}, and returns \texttt{false} otherwise.
    \item \texttt{OR}: Logical operator that combines two or more operator trees. This operator returns \texttt{true} if and only if at least one of the operator trees combined by this operator evaluates to \texttt{true}, and returns \texttt{false} otherwise. 
    \item \texttt{NOT}: Logical operator that negates the output of another the given tree. This operator returns \texttt{true} if and only if the operator tree provided to this operator evaluates to \texttt{false}, and returns \texttt{false} otherwise. 
\end{itemize}

Together, these operators can build complex policies that can later be evaluated to determine whether a client has access to a resource on the blockchain network, or not.

\subsection{Combining Policies}
\begin{figure}[H]
    \centering
    \includegraphics[width=8cm, height=5cm]{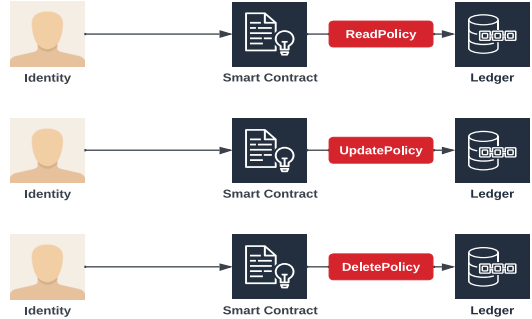}
    \caption{Combining multiple policies. Each smart contract has a different purpose and might need different policies for different operations. Multiple policies can be defined in a single smart contract, and depending on the operation requested by the client, the correct validation policy will be selected and used for access control.} 
    \label{combining-policies}
\end{figure}

As described in the previous subsection, an access policy is a rule that enforces an X.509 certificate to possess a particular combination of attributes and values. These access policies can be used in Hyperledger Fabric to verify whether an entity invoking a smart contract has sufficient permissions to invoke the endpoint. Figure \ref{combining-policies} shows a simplified example of a client invoking three different operations on a smart contract: reading an asset, updating the asset, and deleting the asset. 

As can be seen in the image, the invoked smart contract has a different access policy for each of the three supported operations. For example, a client might be able to satisfy the \texttt{ReadPolicy} with its X.509 certificate, but might not be able to satisfy the \texttt{UpdatePolicy} and \texttt{DeletePolicy}. Therefore, this client will only be allowed to read the asset and will be denied access when it tries to update or delete the asset.

\subsection{Combining ID's}
In Hyperledger Fabric, IDs are composed of X.509 certificates \cite{x509-certificate}, issued by Certificate Authorities and managed by Membership Service Providers. Research into combining multiple such X.509 certificates has not been published to the date of writing. In fact, X.509 certificates cannot be combined by a simple merge, since the X.509 standard \cite{x509-standard} does not allow this. Therefore, for this study, alternative ways of combining multiple X.509 certificates had to be found.

\begin{figure}[H]
    \centering
    \includegraphics[width=9cm, height=7cm]{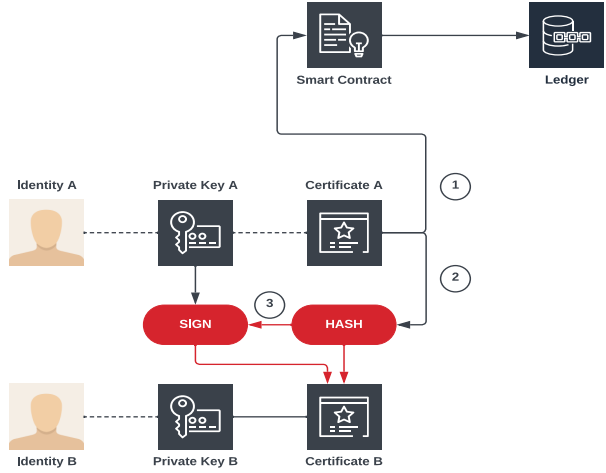}
    \caption{Combining multiple ID's. First, the X.509 certificate of identity A is hashed. Next, this hash is signed with the private key of identity A. Finally, these two values, \texttt{hash(certificate)} and \texttt{sign(hash(certificate))}, are added to the X.509 certificate of identity B as custom attributes.} 
    \label{combining-ids}
\end{figure}

The solution proposed in this study can integrate one X.509 certificate, referred to as the “parent”, into another X.509 certificate. The process by which this integration can be realized is visualized in Figure \ref{combining-ids} and described below. 

\begin{itemize}
    \item First, the member invokes a special smart contract using certificate A (the member is authenticated with certificate A to blockchain network). This smart contract then extracts the certificate from the request, and subsequently stores it into a hashmap on the distributed blockchain ledger;
    \item Second, the member creates the SHA-256 hash of certificate A.
    
    \item Third, the member signs the obtained SHA-256 hash using private key corresponding to certificate A.

    \item Fourth,  the member provides the previously performed hash value and signature to an admin using the \textit{certgen} tool. These arguments allows the \textit{certgen} tool to set the \texttt{hfa.ParentHash} 
    and \texttt{hfa.ParentSignature} attributes of the child certificate (e.g., certificate B).

\end{itemize}

Whenever a client now invokes a smart contract on the blockchain network using identity B, this smart contract can verify that this client also owns identity A, since it needed access to private key A in step (3) to calculate the \texttt{hfa.ParentSignature} attribute. If the client would not have access to this private key, the signature provided in this attribute cannot be valid. Since certificate A was previously stored on the ledger in step (1), the invoked smart contract has access to the public key of identity A, and could therefore easily establish that the provided signature was forged, thus denying access to the network. 

Having established that the client invoking the smart contract with identity B also owns identity A, the smart contract can retrieve the certificate of identity A from the hashmap stored on the distributed ledger, and use it to make access control decisions. The proposed smart contract has been implemented and made available in the GitHub repository\cref{github-repository}. This implementation currently supports one parent certificate to be set in the \texttt{hfa.ParentHash} and \texttt{hfa.ParentSignature} attributes, although it can easily be extended to support multiple parents or recursive ancestor lookups in the future. 

The proposed solution to combine multiple ID's can be particularly useful for decentralized applications and IoT-device applications, where the device or application belongs to a specific owner. In these cases, the identity and access rights of the applications can easily be identified by setting the owner's certificate as the parent certificate within the X.509 certificate of each application. Furthermore, it guarantees that if an application belongs to user B, and therefore contains the hash of user's B identity in its X.509 certificate, it will not be able to access data related to user A.

\subsection{Workflow in a blockchain-IPFS-based network}

\begin{figure}[H]
    \centering
    \includegraphics[width=11cm, height=7.5cm]{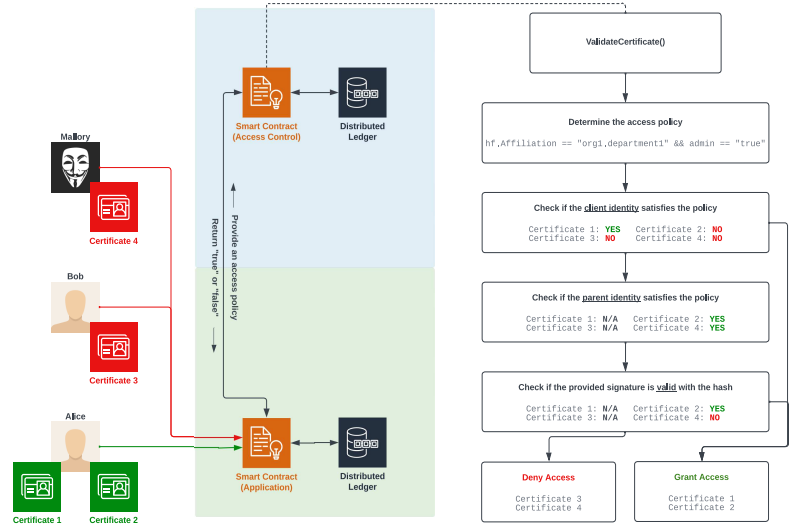}
    \caption{Final system design\cref{github-images}, combining all discussed concepts. Certificate 1 is granted access to the resource since it satisfies the defined access policy. Certificate 2 is granted access to the resource since it contains the \texttt{hfa.ParentHash} and \texttt{hfa.ParentSignature} attributes, which connects it to certificate 1. Certificate 3 is denied access since it does not satisfy the access policy, while certificate 4 is denied access since it contains an invalid hash signature.} 
    \label{combining-final}
\end{figure}

In the previous subsections, the proposed methods of combining multiple ID's, attributes, and policies have been discussed on an individual basis. This subsection will explain how these three concepts will fit together, and how this combined design has been implemented using Hyperledger Fabric. Figure \ref{combining-final} shows a simplified version of the final system architecture\footnote{\label{github-images}More detailed version available at \url{https://github.com/daangordijn/Fabric-Access-Control/blob/master/images}}.

The final system design consists of four main components, which will be described below. 

\textit{Fabric CA Server} A Fabric CA Server instance will be used to issue certificates to various nodes and clients within a particular organization. Fabric CA plays a key role when combining multiple ID's, as it is responsible for creating the basic X.509 certificates and their corresponding private keys, as well as setting the \texttt{hfa.ParentHash} and \texttt{hfa.ParentSignature} attributes if applicable.

\paragraph{Security Smart Contract} The security smart contract is a custom-made smart contract that has two responsibilities. 
\begin{itemize}
    \item First, this smart contract is responsible for maintaining the “parent” X.509 certificates stored on the ledger, as described in Section 5.1. Clients that want to combine two identities, e.g., identity A and identity B, have to invoke this smart contract with identity A. The smart contract will then calculate the SHA-256 hash of the provided certificate, store it in the hashmap on the ledger, and return the hash to the client. Now, the client can calculate the signature and set the required attributes.  

    \item Second, this smart contract can be invoked by other smart contracts that live on the blockchain network to determine whether a client satisfies a particular access policy. Smart contracts can make use of the \texttt{ctx.stub.invokeChaincode()} method to invoke this security smart contract, provide the access policy that has to be validated, and will then be returned a boolean value indicating whether the client certificate satisfies the specified policy. The internal logic of this smart contract method is visualized on the right side in Figure \ref{combining-final}. 
\end{itemize} 

\paragraph{Client Smart Contract(s)} The client smart contracts are basic smart contracts that allow clients of the blockchain network to interact with the ledger. Examples of such smart contracts are the \texttt{asset-transfer} or \texttt{commercial-paper} chaincodes provided in the \texttt{fabric-samples} repository\footnote{\label{fabric-samples}Available at \url{https://github.com/hyperledger/fabric-samples}}. While previously, these smart contracts had to implement their business logic to validate whether a client has access to the requested resource, developers are now able to simply invoke the security smart contract using the \texttt{ctx.stub.invokeChaincode()} method of the Hyperledger Fabric SDK, and use the returned boolean to allow or deny the client from accessing the requested resource. 

\paragraph{Client Application(s)} The client applications are basic applications that allow clients of the blockchain network to more easily interact with smart contracts, instead of having to use the peer CLI. Examples of such client applications are the \texttt{asset-transfer} or \texttt{commercial-paper} applications provided in the \texttt{fabric-samples} repository\cref{fabric-samples}. To client applications, changes made to the proposed solution are not visible, except for the fact that some X.509 certificates containing valid \texttt{hfa.ParentHash} and \texttt{hfa.ParentSignature} attributes will now be granted access, while they would previously have been denied access from the network. 

In summary, this section has presented a solution for combining multiple ID's, attributes, and policies in Hyperledger Fabric. Since this solution can be fully implemented using a single smart contract, the core components of the Hyperledger Fabric blockchain can remain unchanged. In the next section, a performance analysis will be presented, which analyses the increase in runtime due to the invocation and execution of the security smart contract. 

\section{Results}
\label{sec:Results}
One of the most important considerations when proposing a new implementation is to minimize the latency and maximize the transaction throughput. To objectively analyze these performance indicators, two benchmarks of the implemented smart contract were performed with the help of the Hyperledger Caliper \cite{hyperledger-caliper} blockchain benchmarking tool:

\begin{itemize}
    \item \textbf{Basic}: This benchmark analyzes the average latency and throughput when the entity that submits the transaction proposal can satisfy the access policy with its own X.509 attributes; and
    \item \textbf{Parent}: This benchmark analyzes the average latency and throughput when the entity that submits the transaction proposal can only satisfy the access policy with a parent certificate.
\end{itemize}

The exact configuration files that have been used to perform these two benchmarks can be found in the \texttt{caliper} directory of the public GitHub repository\cref{github-repository}. 

\begin{figure*}[btp]
    \centering
    \includegraphics[width=1.0\textwidth]{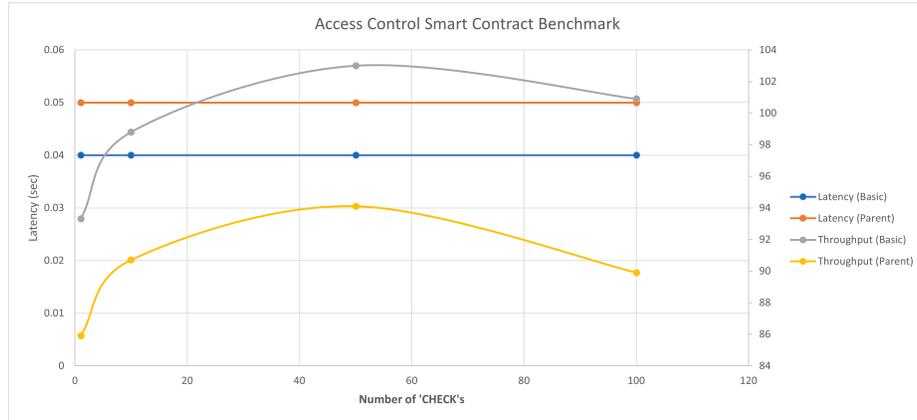}
    \caption{Average latency and throughput of the access control smart contract, measured using the Hyperledger Caliper benchmarking tool. The blue and grey lines respectively show the average latency and throughput that corresponds to the case where the submitting entity satisfies the access policy with its own attributes, while the orange and yellow lines show the case where the access policy had to be satisfied with the parent X.509 certificate, i.e., using the \texttt{hfa.ParentHash} and \texttt{hfa.ParentSignature} attributes.} 
    \label{results-figure}
\end{figure*}

During this study, all benchmarks were performed on a virtual machine running Ubuntu 20.04 LTS, with a total RAM memory of 8GiB. The results that have been obtained are listed in Table \ref{results-table} and visualized in Figure \ref{results-figure}. All reports generated by Hyperledger Caliper can be found in the previously mentioned GitHub repository.

\begin{table}[H]
    \centering
  \begin{tabular}{ |p{1.15cm}||p{1.5cm}|p{1.5cm}||p{2cm}|p{2cm}| }
  \hline
  Checks ($n$) & Latency (Basic) & Latency (Parent) & Throughput (Basic) & Throughput (Parent)\\
  \hline
  1       & 0.04 s& 0.05 s& 93.3  tx/s& 85.9 tx/s\\
  10      & 0.04 s& 0.05 s& 98.8  tx/s& 90.7 tx/s \\
  50      & 0.04 s& 0.05 s& 103.0 tx/s& 94.1 tx/s\\
  100     & 0.04 s& 0.05 s& 100.9 tx/s& 89.9 tx/s\\
  \hline
\end{tabular}
\caption{Average latency and throughput of the access control smart contract, measured using the Hyperledger Caliper benchmarking tool. Each row reports the measured latency and throughput associated with validating the submitted X.509 certificate on the defined access policy, which consists of $n$ attribute checks.}
\label{results-table}
\end{table}

As can be seen in the image, the average latency increases linearly with the number of attribute checks that have to be performed by the smart contract. On the contrary, the average throughput decreases exponentially with this same number of attribute checks. In addition, as can be seen in the image, the performance corresponding to satisfying the access policy with a parent X.509 certificate is slightly worse compared to satisfying this same access policy with its own attributes.

Finally, to objectively quantify these benchmark results, a base case was created and benchmarked using the same Hyperledger Caliper configuration. The smart contract method invoked during this base case benchmark immediately returned a Boolean value, without running any additional code. Hyperledger Caliper reported the average latency of this benchmark to be 0.04 seconds, and the average throughput to be 102.1 transactions per second. Comparing these values with the values listed in Table \ref{results-table}, it can be concluded that the increase in latency and decrease in throughput is very small. When keeping the number of attribute checks below 100, which is considered to be sufficient in most real-world applications, the decrease in performance can be disregarded. 

\section{Discussion}
\label{sec:Discussion}
%This section will discuss the process and technologies that were used to achieve the results and conclusions described in this paper. First, a literature study was conducted on existing literature in the field of secure access control within blockchains, and in particular Hyperledger Fabric. For this, several search queries have been defined, which are included in Appendix A, and executed using Google Scholar. Ultimately, it turned out that there was no existing literature that specifically related to the research topic of combining ID's, attributes, and policies within Hyperledger Fabric.

%Next, a concept solution was defined which could be used to combine these IDs, attributes, and policies. Using Hyperledger Fabric's smart contract technology, these concepts were then implemented and rolled out on a local test network. Subsequently, a demo application and demo smart contract were implemented to analyze the behavior of these concepts. Finally, the implementation was benchmarked using Hyperledger Caliper, after which these results were analyzed and compared to a base case. Based on these results, it was concluded that for real-world scenarios, the performance impact caused by the implemented smart contract is minimal. 

While this paper provides a working solution to solve the identified problem within Hyperledger Fabric, some improvements can be explored in future research. First, although the benchmarks performed by Hyperledger Caliper indicate that the performance impact caused by the proposed implementation is minor, research could be done into ways of improving the algorithms used to validate access policies within the smart contract. Second, the proposed implementation currently only allows users to set one certificate as their parent certificate using the \texttt{hfa.ParentHash} and \texttt{hfa.ParentSignature} attributes. Future research could be done to study whether multiple such parent certificates can be set, for example by allowing array-typed values for these two attributes. Third, since the proposed implementation only allows users to define complex access policies by combining one or more \texttt{EQUALS} or \texttt{INCLUDES} operators using the \texttt{AND}, \texttt{OR}, and \texttt{NOT} operators, research could be done into ways of allowing users to define even richer access policies. Finally, clients must currently store their private key data using file system wallets, which are considered insecure \cite{file-system-wallet}. Future research could be done to allow users to store their private key data in Hardware Security Modules (HSM) to improve the security of this data.

\section{Conclusions}
\label{sec:Conclusion}
%Traditional permissionless blockchain technologies are not sufficient for enterprise-level applications, where privacy and trust are critical. Hyperledger Fabric solves this issue by being a permissioned blockchain technology and using concepts such as identities, channels, and private data collections to create this level of privacy and trust.

One of the major problems of Hyperledger Fabric is that its current access control mechanism is not flexible enough for business scenarios. This study aimed to solve this issue by combining multiple ID's, attributes, and policies with the components that regulate access control.

First, to combine multiple ID's within Hyperledger Fabric, a technique has been proposed that hashes and signs one certificate, referred to as the parent certificate, and adds this hash and signature as attributes to another certificate. A smart contract has been implemented that verifies the ownership of this parent certificate. 

%In addition, this technique stores this full parent certificate in a hashmap on the distributed blockchain ledger. Upon receiving a transaction proposal, smart contracts on the blockchain can retrieve this parent certificate from the distributed ledger and verify ownership using the provided signature. 

Second, to combine multiple attributes, a flexible logic within a smart contract has been proposed that allows access policies to be defined using policy checks combined with Boolean operators. Finally, to combine multiple policies, a technique has been proposed that maintains multiple policy definitions on the distributed ledger, which can dynamically be selected as the validating policy depending on the method invoked with the transaction proposal. 

Finally, in terms of performance, it has been established that for real-world applications the performance impact is negligible. For access policies with less than 100 attributes to check, the increase in average latency is below 0.01 seconds compared to the base case of always allowing access. However, an increase in average latency of 0.1 seconds has been measured when comparing the case where the access policy is satisfied with a member's own attributes with the case where the access policy is satisfied with a member's parent certificate.

%\input{sections/responsible_research}

%\input{appendices/appendix-a}

%\printbibliography

\bibliography{references}
\end{document}